HECTOR GIACOMINI

# Gibbs and Einstein in 1902

## *Parallel Constructions in Statistical Mechanics*

▼ **ABSTRACT** This article reexamines the striking conceptual and formal parallels between Josiah Willard Gibbs's *Elementary Principles in Statistical Mechanics* and Albert Einstein's first paper on statistical mechanics. Although their derivations follow different paths, both arrive at the canonical distribution and construct closely related theoretical frameworks for thermal equilibrium. Each offers a mechanical grounding of the second law of thermodynamics and defines temperature as the quantity that equalises when systems exchange energy, thereby characterising equilibrium without recourse to kinetic theory. In both cases, the constructions rely on general mechanical principles—most notably energy conservation and Liouville's theorem—combined with probabilistic reasoning, and deliberately set aside Boltzmann's kinetic approach. Whereas Gibbs's treatise quickly became a canonical reference, Einstein's 1902 paper remained marginal in both physics and its historiography for more than half a century. Focusing primarily on that first article, while also considering the closely related papers of 1903 and 1904, the present study reviews the main historiographical interpretations of their reception. Particular attention is given to Paul Hertz's two critical articles of 1910, which identified a close affinity between Gibbs's and Einstein's approaches and played an important role in Einstein's subsequent distancing from his early statistical programme, following a private meeting in Basel that Einstein sought in preference to a public controversy. Drawing on new archival evidence concerning the rapid European dissemination of Gibbs's book in the spring of 1902, the article reassesses the plausibility of Einstein's later claim that he was unaware of Gibbs's work at the time. It also highlights Max Planck's extensive 1903 review, published in the influential *Beiblätter zu den*

**Hector Giacomini** 0009-0007-9194-7193 • Institut Denis Poisson, Tours, France, correspondence: Hector Giacomini, Institut Denis Poisson (Université d'Orléans, Université de Tours, CNRS), 37200 Tours, France, giacominihector@gmail.com

*Annalen der Physik*, which offered a particularly clear physical discussion of statistical fluctuations within the Gibbsian framework.

## Introduction

Albert Einstein's first papers on statistical mechanics, published between 1902 and 1904, occupy an unusual place in the history of modern physics. They form a coherent and ambitious attempt to derive thermodynamic equilibrium from general mechanical and probabilistic principles, yet they remained marginal in the subsequent development and historiography of statistical mechanics. By contrast, Josiah Willard Gibbs's *Elementary Principles in Statistical Mechanics*, published in New York and London in 1902, rapidly became the canonical reference for the ensemble formulation of the field. The proximity between Gibbs's treatise and Einstein's first statistical paper has often been noted. Both authors introduce stationary probability distributions, rely on energy conservation and Liouville's theorem, and arrive at forms of the canonical distribution that would later become central to equilibrium statistical mechanics. Yet Einstein did not cite Gibbs in his papers of 1902–1904, and later stated that he had been unaware of Gibbs's work at the time.

This situation has usually been interpreted as a case of independent reconstruction. Einstein's early statistical papers have been read as emerging primarily from his engagement with Boltzmann's writings and from his own attempt to ground thermodynamics in mechanics. His marginal institutional position in 1902, outside the university system and without a stable academic post, seemed to reinforce the assumption that Gibbs's newly published treatise was unlikely to have played any role in the genesis of his first paper on statistical mechanics. At the same time, the later reception of Einstein's early statistical work contributed to this interpretation: the papers were retrospectively integrated into the continuity of Einstein's intellectual development, but they did not become foundational reference points for statistical mechanics itself.

The present article reopens this question by combining textual comparison, reception history, correspondence, and documentary evidence concerning the circulation of Gibbs's book. It does not claim to prove that Einstein read Gibbs's *Elementary Principles* before writing his first statistical paper. Such a conclusion is not supported by the surviving evidence. Its aim is rather to show that the usual assumption that Gibbs's book was not materially accessible to Einstein is no longer tenable, and that Einstein's statistical papers of 1902–1904 must be situated within a documentary environment in which Gibbs's work had already begun to circulate rapidly in Europe. Recent archival evidence shows that copies of Gibbs's treatise reached several major European university libraries in the spring of 1902, including the University Library of Basel on April 10, more than 2 months before Einstein submitted his first statistical paper to the *Annalen der Physik*. This does not settle the question of influence, but it changes the conditions under which that question must be posed.

The article first reconstructs the conceptual structure of Gibbs's *Elementary Principles* and the rapid dissemination of the book in Europe. It then examines Einstein's situation in the spring of 1902: his professional uncertainty, the gaps in the surviving correspondence, his channels of scientific exchange, and the abrupt transition from his April 1902 paper on electrochemistry to the June 1902 paper on statistical mechanics. The core of the article is devoted to Einstein's first statistical paper, whose formal architecture is compared with Gibbs's ensemble-based framework. The papers of 1903 and 1904 are also considered, as extensions of the same programme. The 1903 paper generalises the dynamical framework in an even more abstract direction, while the 1904 paper introduces fluctuations in a context in which Gibbs's treatment, and especially Planck's 1903 review of it, had already made the relevant conceptual tools publicly available in the German-speaking literature.

A central part of the argument concerns reception. Einstein's 1902 paper did not become a standard point of reference in statistical mechanics. It was known, but it was soon classified as compatible with a framework more fully and systematically represented by Gibbs. This reception history is especially visible in the 1911 review article by Paul and Tatiana Ehrenfest, where Gibbs occupies a central place whereas Einstein's early papers are treated only briefly. The most revealing episode, however, is Paul Hertz's public critique of 1910. Hertz explicitly compared Gibbs's and Einstein's approaches in the *Annalen der Physik* and argued that the fundamental ideas and formulae in this domain were already contained in Gibbs's work. Einstein did not answer this challenge through a public controversy. Instead, he sought a private discussion with Hertz, arranged a meeting in Basel, and then published a short note in 1911 acknowledging that, had he known Gibbs's treatise at the time, he would not have published his papers of 1902–1904. This episode is important because it shows that the Gibbs–Einstein parallel was not merely a later historiographical construction: it had already been clearly perceived by an informed contemporary and elicited a cautious response from Einstein himself.

The final sections draw out the broader historiographical implications of this case. The issue is not one of priority in a narrow sense, nor does it require replacing a narrative of independence with a simple claim of borrowing. Rather, the case invites a reconsideration of how historians interpret silence, citation practices, material accessibility, and retrospective self-testimony. Einstein's later statements about Gibbs belong to a wider pattern in which he often presented theoretical physics as an essentially solitary enterprise, governed by intuition and conceptual necessity rather than by visible dependence on texts, networks, and prior programmes. The possibility of early access to Gibbs's treatise therefore matters not only for the history of statistical mechanics, but also for the historiography of Einstein's scientific self-image and for the broader question of how individual creativity is reconstructed within collective scientific traditions.

## Josiah Willard Gibbs's Elementary Principles in Statistical Mechanics

Published in New York and London in March 1902, *Elementary Principles in Statistical Mechanics* represents the most mature synthesis of Josiah Willard Gibbs's thermodynamic and mechanical thought.[1] At the time of the book's publication, Gibbs was already a well-established and highly respected figure among European physicists, owing to his fundamental contributions to thermodynamics in the 1870s and 1880s. His work on the graphical and analytical formulation of thermodynamic potentials had been widely recognised and disseminated in Europe, notably through the writings of James Clerk Maxwell and the energeticist tradition associated with Wilhelm Ostwald. Gibbs was thus far from an obscure or marginal author in 1902. What was new, and largely unprecedented, was not his authority as a thermodynamicist, but the systematic extension of his approach to the foundations of statistical mechanics in a single, unified treatise.

The book appeared in the commemorative series of the *Yale Bicentennial Publications* (1901–1903), which comprised 24 volumes. It was the product of many years of reflection, written in the final years of Gibbs's life and published shortly before his death in April 1903.

The treatise introduces, in a systematic way, the method of statistical ensembles: microcanonical, canonical, and grand canonical. Rather than analysing an isolated mechanical system, Gibbs considers an ideal collection of identical systems subject to the same constraints, and applies probabilistic reasoning to deduce their average behaviour. Gibbs does not attempt to explain how thermal equilibrium is reached; instead, he characterises the probability distributions that define it. Hamiltonian mechanics appears only to establish Liouville's theorem and the conservation of total energy. Apart from these two properties, the entire statistical structure rests on general probabilistic principles, independent of dynamical details.

Gibbs assumes that among all constants of motion (energy, momentum, angular momentum, and so on), only energy retains a macroscopic role in the description of equilibrium. In other words, when one passes from a microscopic description governed by the equations of motion to a global description based on mean quantities such as temperature or pressure, energy remains the sole constraint determining the distribution of microstates. This assumption justifies why, in Gibbs's ensembles, the distribution depends only on energy. This property is essential: without it, the distribution would not be a function of energy alone, and equilibrium could not be represented by the microcanonical ensemble.

Gibbs derives the exponential law of the canonical distribution by a simple and elegant factorisation argument: for two independent systems, the joint probability density must factorise, which imposes an exponential dependence on energy.

The work is characterised by its formal rigour and high degree of abstraction. Far from the mechanical analogies still present in Boltzmann's writings, Gibbs proposes a mathematical construction based on ensembles of microstates and probability

1 Gibbs (1902).

functions, without resorting to concrete particle models. The aim is no longer to retrace the formation of equilibrium, but to formulate the logical conditions that define it. In this sense, Gibbs grants statistical mechanics an autonomous status, distinct from both kinetic gas theory and phenomenological thermodynamics.

The book is often regarded as a landmark in the history of equilibrium statistical mechanics, not because it created the field ex nihilo, but because it provided a systematic, unified, and highly abstract formulation of concepts and methods that had been developed over previous decades. By recasting statistical mechanics in terms of ensembles and probability distributions defined on phase space, Gibbs offered a formal language and logical structure that proved particularly well suited to generalisation and long-term use throughout the 20th century.

In contrast to Boltzmann's historically earlier work, which combined probabilistic reasoning with dynamical considerations and explicit discussions of time evolution, Gibbs deliberately set aside questions of dynamical approach to equilibrium. His focus lay instead on the definition of equilibrium states as stationary probability distributions, characterised by general constraints such as energy conservation. While Boltzmann had already employed ensemble-like constructions and probabilistic arguments, Gibbs elevated these tools to the status of a self-contained theoretical framework, largely independent of specific mechanical models. Despite these differences in emphasis and presentation, both authors shared the same fundamental aim: to understand how thermodynamic laws can be grounded in the principles of mechanics.

For detailed historical and conceptual analyses of Gibbs's work, see works by Klein (1987) and Darrigol and Renn (2013).[2] These studies show how the Gibbsian formalism became the foundational reference of modern statistical physics.

### *Dissemination*

Recent archival research has brought to light a mode of dissemination of Gibbs's *Elementary Principles* that has been overlooked in the historiography. As part of an international donation campaign organised by Yale University for its bicentennial, Gibbs's book was distributed to numerous European universities as early as the spring of 1902. The operation was coordinated by the New York publisher Charles Scribner's Sons and formed part of the institutional series of the *Yale Bicentennial Publications*, a collection of 24 volumes intended to represent Yale University's scientific output.

Acquisition registers from major European university libraries—including Oxford, Göttingen, Heidelberg, Basel, Vienna, and Strasbourg—show that the Gibbs volume arrived with remarkable rapidity, typically between April 1 and April 19, 1902. In parallel, documentary evidence indicates that by mid-March more than 40 individual scientists and several learned societies had already received copies of the book directly.

2 Klein (1987); Darrigol & Renn (2013).

This pattern of institutional distribution establishes that Gibbs's treatise was materially accessible in Europe by April 1902. The speed, scale, and organised character of this dissemination have not been previously taken into account in assessments of Einstein's early statistical work.[3]

The copy received by the University Library of Basel on April 10, 1902, was the first to reach continental Europe, and remains the only surviving 1902 edition currently held in Switzerland. At the beginning of the 20th century, the University Library of Basel occupied a singular position in the Swiss documentary landscape. Founded in 1471, it is one of the oldest university libraries in Europe. Contrary to the image of an institution oriented mainly towards the humanities and theology—a reflection of its long humanist tradition linked to Erasmus of Rotterdam and the reputation of its faculty of philosophy—the library had, since the 19th century, built a substantial scientific collection by Swiss standards.

This scientific orientation was grounded in a much older intellectual tradition. Since the 17th and 18th centuries, Basel had been home to some of the greatest mathematicians in Europe, including Jacob Bernoulli, Johann Bernoulli, Daniel Bernoulli, and Leonhard Euler, all of whose early academic careers unfolded at the University of Basel.

At the same time, Basel's faculty of philosophy enjoyed a European reputation, reinforced by the presence of Friedrich Nietzsche, who taught classical philology at the University of Basel from 1869 to 1879.

## Albert Einstein in 1902

### *Einstein's Personal Situation in Early 1902*

In September 1901, Albert Einstein was hired as a private tutor by Jakob Nüesch, the director of an educational institution in Schaffhausen, about 50 km from Zurich. This non-academic position, probably based on a private agreement, lasted about four and a half months, until the end of January 1902.

In a letter to his friend Conrad Habicht, dated February 4, 1902, Einstein mentioned a disagreement with Nüesch and stated that he had left the position abruptly.[4] Several biographers believe that the post had been intended to last a full year but was terminated prematurely. At that moment, Einstein was going through a period of professional and personal uncertainty: all his applications for university assistantships had failed, his doctoral thesis submitted in Zurich had been withdrawn on February 1, 1902, and his family was facing financial difficulties in Milan. Mileva Marić, who had returned to Serbia, had just given birth to their daughter Lieserl in great secrecy.

3 A detailed reconstruction of this network of distribution is presented in Giacomini (2026).

4 A. Einstein to C. Habicht [Letter] (1902, Feb. 4), CPAE 1 English, doc. 133. CPAE refers to *The Collected Papers of Albert Einstein*; references are arranged by volume, language, and document number. CPAE 1 German is Stachel et al. (1987a); CPAE 1 English is Stachel et al. (1987b); CPAE 5 English is Klein et al. (1995).

Arriving in Bern on February 2 or 3, 1902, Einstein rented a modest room and survived thanks to private lessons in physics and mathematics. Only on June 23 did he finally obtain a stable position at the Swiss Federal Office for Intellectual Property, after several months of waiting and uncertainty.

Max Talmey, a former acquaintance of Einstein's family in Munich, then living in New York, visited Einstein in Bern in April 1902 and had presumably left New York for Europe sometime in the preceding weeks. Talmey (born Talmud) had regularly lent the young Einstein scientific and philosophical books between 1889 and 1894, when he was a medical student. After emigrating to New York in 1895, he settled as an ophthalmologist in Manhattan.

In a book published in 1932, Talmey recounts this visit: he found Einstein "living in a small, poorly furnished room," absorbed in theoretical reflections yet struggling with material difficulties.[5] Einstein told him that his career had been hindered by "obstacles placed by jealous people" and handed him an offprint of his only published paper at the moment, on capillarity phenomena.

Talmey's taste for reading and the circulation of ideas had already been evident during his years in Munich. A curious and polyglot student, he devoted much of his free time to reading scientific and philosophical treatises, which he readily lent to the young Einstein. After settling in New York, he remained an avid reader of recent publications and likely frequented the major academic bookshops of Manhattan, a pattern of intellectual mediation that has been further documented and contextualised in a recent article.[6]

### *Correspondence Gaps and Channels of Scientific Exchange*

From early April 1902 onward, the documentation concerning Einstein becomes markedly sparse. As indicated by the editorial apparatus of the *Collected Papers*, only two letters from the second half of 1902 are preserved.[7] The surviving correspondence thus provides no direct evidence concerning the conceptual genesis of his first two papers on statistical mechanics.

A short note addressed to Conrad Habicht in April 1902 is particularly revealing. Einstein reports having received a message from his former classmate Jakob Ehrat, an assistant at the Swiss Federal Polytechnic in Zurich. Ehrat asked him to remind Habicht to return a mathematics book he had borrowed and that belonged to the Polytechnic library.[8] A former fellow student of Einstein, Ehrat—like Habicht, originally from Schaffhausen—already acted as a link between Zurich and Einstein, who had recently settled in Bern.

---

5 Talmey (1932).
6 Gordin (2022).
7 A. Einstein to M. Marić [Letter] (1902, Jun. 28 or later), CPAE 5 English, doc. 1; A. Einstein to H. Wohlwend [Letter] (1902, Aug. 15–Oct. 3), CPAE 5 English, doc. 2.
8 A. Einstein to C. Habicht [Letter] (1902, Apr., date uncertain), CPAE 1 English, doc. 139.

This brief letter reveals the existence of a documentary network connecting Zurich and Bern. A few months earlier, in a letter to Mileva Marić dated December 28, 1901, Einstein had written that he wanted to study what Lorentz and Drude had written on the electrodynamics of moving bodies, adding that Ehrat would have to obtain the literature for him.[9] The April 1902 message shows that this channel was still functioning: Einstein and Habicht continued to rely on Ehrat to access books and articles from the Polytechnic library, maintaining a flow of scientific resources despite their geographical dispersion.

Only two written traces testify to Einstein's scientific activity during this interval: his first paper on statistical mechanics, received by the *Annalen der Physik* on June 26, 1902, and the second, received on January 26, 1903.[10] The first attempts to derive the principles of thermodynamics from the laws of mechanics, while the second deepens and generalises this approach, extending it to more complex systems. These two closely connected texts constitute the only direct windows onto Einstein's intellectual activity in the spring of 1902 through January 1903.

It is generally recognised that the surviving correspondence between Albert Einstein and Mileva Marić is fragmentary. Of the known letters from the years 1897–1903, 54 are preserved: 43 from Einstein to Marić and 11 from Marić to Einstein. This surviving exchange is substantial, but it is also highly uneven. In particular, Marić's side of the correspondence is only sparsely represented, and the gaps are especially significant for the period around 1901–1903.

The fragmentary character of the corpus has often been discussed in connection with the birth of Einstein and Marić's daughter Lieserl in early 1902. Yet several letters explicitly mentioning the child have survived. This makes it difficult to treat the surviving pattern of the correspondence as the straightforward result of a systematic suppression of all references to Lieserl. More cautiously, the correspondence should be regarded as an incomplete and unevenly preserved record, shaped by circumstances that remain only partly recoverable.

For the present argument, the essential point is not to identify the precise cause of the loss, but to register its historiographical consequence. The missing letters may well have contained discussions of Einstein's scientific work, his readings, and his exchanges with Marić during a formative period. Their absence therefore deprives historians of any direct epistolary evidence concerning Einstein's readings and intellectual influences during the spring and summer of 1902.

### *Einstein's Early Scientific Work Before June 1902*

It is useful to recall the first two papers Einstein published in the *Annalen der Physik*. They reflect a training still largely empirical and a desire to enter the scientific community, while already showing his interest in the relationship between microscopic phenomena and macroscopic laws.

9 A. Einstein to M. Marić [Letter] (1901, Dec. 28), CPAE 1 English, doc. 131.
10 Einstein (1902a; 1903).

(1) Einstein's first paper, entitled "Conclusions Drawn from the Phenomena of Capillarity," appeared in March 1901.[11] It deals with the molecular forces responsible for surface tension in liquids and seeks to relate observable macroscopic effects to attractive interactions between molecules. The approach remains classical, but it already reveals Einstein's tendency to search for microscopic foundations of phenomenological laws. This paper was written in Zurich shortly after his graduation from the Swiss Federal Polytechnic in July 1900.

(2) The second article, "On the Thermodynamic Theory of the Difference in Potential Between Metals and the Completely Dissociated Solutions of Their Salts," was received by the *Annalen der Physik* on April 30, 1902.[12] Einstein wrote it during his first months in Bern. The study applies the principles of classical thermodynamics to interpret electrochemical phenomena and shows Einstein's growing mastery of physical reasoning, although it remains entirely rooted in the macroscopic framework of thermodynamic equilibrium.

Einstein later regarded these two articles as modest preparatory work. He referred to his initial theoretical attempts as mere exercises, downplaying their lasting significance. Yet these first publications mark the beginning of a more ambitious phase of reflection: from the spring of 1902 onward, Einstein turned for the first time toward a probabilistic formulation of the laws of thermodynamics.

## Einstein's 1902 Paper on Statistical Mechanics

At the end of June 1902, Einstein submitted to the *Annalen der Physik* his first paper on statistical mechanics, received on June 26.[13] This text marks his entry into the field: for the first time, he attempted to ground the theory of thermal equilibrium on general mechanical and probabilistic principles. Despite its reference to a "kinetic theory of thermal equilibrium," the approach adopted in the paper differs fundamentally from kinetic theory in the sense developed by Boltzmann.

Einstein begins by postulating the existence of a stationary probability distribution depending only on the total energy of the system. He does not attempt to describe any relaxation process or dynamical approach to equilibrium. Equilibrium is introduced as an invariant statistical state, defined by a probability density constant along mechanical trajectories. In this respect, Einstein's construction is structurally analogous to that of Gibbs, who likewise analysed equilibrium by considering ensembles of mechanically equivalent systems rather than the temporal evolution of a single system.

The strategy of analysing a subsystem in thermal contact with a large system of fixed total energy had already been introduced by Boltzmann as a way of justifying equilibrium distributions through the counting of accessible states. Gibbs was fully

11 Einstein (1901).
12 Einstein (1902b).
13 Einstein (1902a).

aware of this type of reasoning and discusses the situation of a system in contact with a heat reservoir in his treatise, although he chose to ground the canonical distribution primarily in a condition of factorisation for independent systems. Einstein, by contrast, derives the canonical form by starting explicitly from a microcanonical description of the total system and treating the subsystem as embedded in a large reservoir whose number of accessible states varies with energy exchange. While the resulting distribution coincides with Gibbs's canonical form, the reservoir argument plays a more explicit and central role in Einstein's derivation.

From the outset, Einstein presents his work as addressing a problem left open by Maxwell and Boltzmann: the derivation of the laws of equilibrium and of the second law of thermodynamics from mechanics alone. Yet he deliberately abandons Boltzmann's dynamical route. No equations of motion appear in the paper, nor is there any discussion of particle collisions, trajectories, or temporal irreversibility. Equilibrium is not treated as the outcome of an evolutionary process, but as a stationary property of a statistical ensemble.

The paper thus reflects a decisive change of perspective. Whereas Boltzmann sought to explain equilibrium through dynamical evolution, Einstein adopts a strictly formal approach: he assumes an invariant distribution, deduces the relative probabilities of accessible states, and reconstructs thermodynamic quantities from this structure. This move marks a clear break with the kinetic tradition and brings Einstein's analysis remarkably close to Gibbs's ensemble-based framework.

Einstein explicitly refers to Boltzmann for the invariance of the statistical density and the mechanical interpretation of the parameter h. These references concern specific steps in the argument, without explicitly attributing the overall construction to Boltzmann.

Conceptually, the parallels between Einstein's construction and Gibbs's theory are numerous: the hypothesis of a stationary density, the use of a thermal reservoir, the exponential form of the equilibrium distribution, and the reconstruction of thermodynamic quantities from this structure. Despite differences of terminology and presentation, the two texts articulate the same theoretical architecture, which would later become the foundation of equilibrium statistical mechanics. Einstein did not yet systematise this framework as comprehensively as Gibbs, but its essential principles are already present in a strikingly similar form.

In this sense, the 1902 paper marks both the starting point of Einstein's reflection on statistical equilibrium and a moment of strong convergence with Gibbs's structural perspective, combined with a clear departure from Boltzmann's dynamical programme.

### *From Electrochemistry to Statistical Mechanics*

Between April and June 1902, Einstein's intellectual trajectory underwent a striking inflection. Within a span of 8 weeks, he moved from a paper on an electrochemical problem, received by the *Annalen der Physik* on April 30, 1902, to a foundational paper on statistical mechanics, received on June 26.[14]

The contrast between these two texts is pronounced. The earlier work belongs to physical chemistry and addresses concrete electrolytic phenomena within the established thermodynamics of chemical reactions. The later paper, by contrast, develops an abstract probabilistic framework for thermal equilibrium, largely detached from specific physical systems.

This rapid transition from an applied domain to a foundational theoretical construction marks a profound change of perspective. In a matter of weeks, Einstein shifted from the analysis of particular phenomena to the formulation of a general statistical framework for thermodynamics. No surviving document explicitly prepares this reorientation. The transition thus reflects a sudden conceptual consolidation rather than the gradual unfolding of a long-term research programme.

## Historiographical Assessments of Einstein's 1902 First Paper on Statistical Mechanics

Einstein's first paper on statistical mechanics has long occupied an ambiguous position in the historiography. On the one hand, historians have repeatedly emphasised the striking formal proximity between Einstein's 1902 construction and Gibbs's ensemble-based framework; on the other hand, this proximity has almost systematically been interpreted as the result of an independent reconstruction, grounded primarily in Einstein's engagement with Boltzmann's writings. A recurrent tension thus runs through the literature between the recognition of structural similarity and the persistence of an independence narrative. This narrative has generally been maintained without close attention to the material circulation of Gibbs's book in Europe, to the unusually compressed chronology of spring 1902, or to the precarious professional situation in which Einstein produced his first statistical paper.

From the mid-20th century onward, commentators repeatedly noted the close formal parallels between Einstein's 1902 paper and Gibbs's *Elementary Principles*. The editorial introduction to Einstein's statistical papers in the *Collected Papers of Albert Einstein* already remarked, in a neutral manner, that Einstein derived canonical distributions and equilibrium properties in a way closely resembling Gibbs's approach, despite making no reference to the latter.[15] In a more emphatic register, Max Born described Einstein's work as a re-derivation of essential features of statistical mechanics and stressed the near-identity of Einstein's method with Gibbs's canonical formalism, while nonetheless interpreting the convergence as an independent achievement.[16]

A similar pattern characterises classic historical surveys. Martin J. Klein and Jagdish Mehra both underlined the breadth and sophistication of Einstein's early statistical work while repeatedly noting that many of its core elements were already present in Gibbs's book. Yet both authors framed Einstein's construction as emerging primarily from a Boltzmannian tradition and treated the Gibbs–Einstein parallels as cases of independent convergence.[17] The editors of the *Collected Papers of Albert Einstein* followed the same interpretive line: they acknowledged the formal proximity between Einstein's canonical derivations and Gibbs's framework but judged independence more likely, given the absence—at the time—of documentary evidence that Einstein had access to Gibbs's book in 1902.[18] More recent syntheses have largely preserved this assessment. Anne J. Kox, as well as Janssen and Lehner in the *Cambridge Companion to Einstein*, explicitly note that several of Einstein's results had been anticipated by Boltzmann and Gibbs, while reiterating that Einstein appears to have reconstructed the canonical formalism on Boltzmannian grounds alone, in the absence of evidence for early access to Gibbs.[19]

---

14 Einstein (1902b).
15 Stachel et al. (1989, pp. 41–55).
16 Born (1956).

A second strand of historiography has shifted attention from questions of influence to differences of interpretation. Navarro, for example, argues that Gibbs's ensembles rely on an objective notion of probability defined over fictitious collections of mechanically identical systems, whereas Einstein's probabilities should be read in epistemic terms, as expressing incomplete knowledge of an individual system.[20] Darrigol develops related contrasts, emphasising that conceptual interpretations of probability may diverge even when mathematical architectures closely coincide.[21] In these analyses, the proximity between Gibbs and Einstein is fully acknowledged at the level of formal construction, while interpretive differences are foregrounded.

Despite their diversity, these historiographical positions share two robust assumptions: first, that the structural parallels between Gibbs's book and Einstein's 1902 paper are substantial; second, that Einstein's construction is best understood as an independent development grounded in his reading of Boltzmann. The inference from Einstein's silence to his ignorance of Gibbs, however, is typically made without any examination of the material circulation and accessibility of Gibbs's treatise in Europe, and without confronting the unusually compressed chronology under which Einstein's first statistical paper was produced.

Alongside these Gibbs–Einstein comparisons, a number of studies have emphasised the internal coherence of Einstein's statistical trilogy and its continuity with his later work. On this reading, the papers of 1902–1904 constitute the opening phase of a distinctive programme aimed at deriving thermodynamics from very general mechanical considerations through probabilistic reasoning. Reconstructions by Renn and Janssen, for instance, stress the methodological continuity of Einstein's early statistical thinking and its later extension to fluctuations and Brownian motion.[22] While these Einstein-centred readings compellingly illuminate the ambition and coherence

17 Klein (1963; 1967); Mehra (1975).
18 Stachel et al. (1989, pp. 41–55).
19 Kox (2014); Janssen & Lehner (2014).
20 Navarro (1998).
21 Darrigol (2004); Darrigol & Renn (2013).
22 Renn (2005); Duncan & Janssen (2019).

of the trilogy, they generally preserve the conventional assumption of independence from Gibbs, again without analysing the concrete conditions of access to the 1902 treatise.

Finally, many historiographical reconstructions risk introducing a teleological perspective into the reading of Einstein's first paper. The 1902 text is often retrospectively interpreted as the initial step of a long-term programme oriented toward atomism or toward the systematic exploitation of fluctuations. Yet such intentions are not documented in the paper itself. Einstein's 1902 work develops a highly abstract and continuous probabilistic formalism, without explicit commitment to atomic hypotheses, without detailed dynamical modelling, and without any analysis of fluctuations, which would only enter his work 2 years later. The coherence later discerned across the statistical trilogy therefore reflects, at least in part, a retrospective reconstruction informed by Einstein's subsequent achievements.

Over time, the claim of Einstein's independence from Gibbs has acquired the status of an implicit historiographical premise. In a substantial body of influential work, independence is not argued for but stated as an evident fact, sometimes already at the level of abstracts or introductory framing. A representative illustration of this stabilised reading is provided by Gearhart's pedagogical survey of Einstein's early statistical papers.[23] Situating Einstein firmly within a Boltzmannian lineage, Gearhart treats the close parallels between Einstein's 1902 construction and Gibbs's ensemble formalism as unproblematic cases of independent convergence, requiring no special explanation. In this account, Boltzmann's influence is taken as self-evident, while the question of Einstein's possible access to Gibbs's *Elementary Principles* is left unexamined.

A further difficulty arises from the limited scope of Einstein’s explicit references to Boltzmann. This limited acknowledgment poses a historiographical problem for accounts that treat Einstein's 1902 construction as deriving almost entirely from Boltzmann's results. Given Boltzmann's exceptional standing in European physics around 1900, it is not obvious that a young and professionally precarious author could have relied so heavily on these works without explicitly acknowledging such a broad dependence on them. The present study does not seek to replace this narrative with an assertion of direct influence. Rather, it aims to reopen a question that has too often been treated as settled in advance, and to show that the assumption of independence itself requires historical justification—one that must take into account material circulation, accessibility, and the concrete conditions under which Einstein's early statistical work was produced.

## Reception and Citations of Einstein's Paper

Einstein's 1902 paper did not become a point of reference in the subsequent development of statistical mechanics. Although it was known to contemporaries, it

23 Gearhart (1990).

was rapidly classified as conceptually non-fundamental within a field whose general framework was soon dominated by Gibbs's formalism. As a result, the paper exerted little direct influence on later developments and was progressively marginalised in canonical accounts of statistical mechanics.

Between 1902 and 1909, the paper appears to have received no explicit discussion in the literature. Einstein himself referred back to his early statistical work only indirectly. In his 1905 paper on Brownian motion, he alluded to previous analyses without explicitly citing the 1902 text.

A particularly revealing case is provided by the influential 1911 review article on statistical mechanics by Paul and Tatiana Ehrenfest.[24] In this survey, the authors devoted extensive analysis to Gibbs's work, which they treated as the definitive conceptual foundation of the subject. Einstein's papers of 1902 and 1903 are explicitly listed in the bibliography but they receive no substantive historiographical discussion. This extreme brevity cannot be attributed to ignorance; rather, it reflects a historiographical judgment according to which these papers were regarded as compatible reformulations of an already established framework, not as parallel or independent foundational contributions. Significantly, Einstein's 1904 paper on energy fluctuations is entirely absent from the Ehrenfests' review, which moves directly from the equilibrium papers of 1902–1903 to the theory of Brownian motion developed in 1905.

In the following decades, major treatises on statistical physics—most notably that of Richard Tolman—fully adopted the Gibbsian framework, thereby consolidating the marginal status of Einstein's early statistical work.[25] Until the early 1960s, the 1902 paper remained absent from standard textbooks and from most historical accounts of physics, which focused instead on the Einstein of 1905 and later.

From the 1960s onward, historical studies began to reintegrate the 1902 paper into the continuity of Einstein's intellectual development, emphasising its coherence and systematic ambition. This reappraisal, however, was essentially historiographical. It did not reflect a delayed scientific influence, but rather a renewed effort to reconstruct the internal logic of Einstein's early work in statistical mechanics. The reception history of the 1902 paper thus illustrates not simple neglect, but a process of early classification and narrative marginalization, shaped by the rapid consolidation of the Gibbsian formalism and by Einstein's own subsequent disengagement from his statistical papers.

## Context of 1902–1903: The Diffusion of Gibbs's Book

At the time when Einstein was writing his two new papers on the subject, published in 1903 and 1904, Gibbs's treatise was beginning to be known in several European scientific centres. From the spring of 1902 onward, numerous university institutions

24 Ehrenfest & Ehrenfest (1911).
25 Tolman (1938).

and several scientific journals received copies of the volume directly from New Haven, sent by Charles Scribner's Sons on behalf of Yale University as part of the *Yale Bicentennial Publications* series. The New York publishing house thus sent copies to various European journals, among them the *Physikalische Zeitschrift* and *Nature*.[26] Between July 1902 and December 1903, Gibbs's *Elementary Principles* was reviewed on at least four distinct occasions in major European scientific venues, by G. H. Bryan, Max Abraham, Max Planck, and Stefan Meyer.

Max Abraham's review, received by the journal on July 15, 1902, appears to have been the first review published in continental Europe.[27] It was preceded by G. H. Bryan's review in Britain. He clearly presented the structure of the microcanonical and canonical ensembles and emphasised the conceptual significance of the treatise. Earlier that month, on July 1 and 8, 1902, Abraham also presented Gibbs's book in two successive meetings of the Mathematische Gesellschaft zu Göttingen. The *Jahresbericht der Deutschen Mathematiker-Vereinigung* records that Abraham began a "larger report" on Gibbs's *Elementary Principles* on July 1 and "continued his report" on July 8, followed by a discussion focused largely on the status of entropy as a monotonically increasing function.[28] This notice shows that Gibbs's treatise was already being publicly discussed in the German-speaking scholarly community in early July 1902, only a few months after its publication in New York and London. On July 24, 1902, G. H. Bryan published in *Nature* a review highlighting the methodological importance of the treatise and its contribution to a rigorous foundation of statistical mechanics.[29]

### *Max Planck's 1903 Review of Gibbs and the Discussion of Statistical Fluctuations*

In his six-page review published in the *Beiblätter zu den Annalen der Physik* in 1903, Max Planck praised Gibbs's book as the most rigorous and systematic exposition then available of the mechanical foundations of thermodynamics.[30] He emphasised the conceptual unity introduced by the ensemble method, which replaces the dynamical study of a single system with that of an infinite collection of identical systems populating phase space. For Planck, this conceptual shift is both simple and profoundly fruitful.

One of the most striking features of Planck's review is his emphasis on what we would now call *fluctuations*. Planck draws attention to the "anomalies"—the deviations of mechanical quantities from their mean values—and stresses that these anomalies become negligible when the number of degrees of freedom of a system is very large. In his account, the apparent determinism of macroscopic systems is thus explained as the practical disappearance, for an imperfect observer, of the random

26 Giacomini (2026).
27 Abraham (1902).
28 Gutzmer (1902).
29 Bryan (1902).
30 Planck (1903).

properties that still characterise individual microscopic systems. Planck illustrates this point with an example: a body immersed in a thermostat *appears* to possess a well-defined energy, even though, from the statistical point of view, its energy actually fluctuates around a mean value. Although Gibbs had introduced the mathematical apparatus for variances and canonical averages, he had not spelt out their physical implications with the clarity and transparency found in Planck's exposition. In this respect, Planck's review does more than summarise: it provides a particularly clear physical interpretation of Gibbs's theory of fluctuations, showing how macroscopic stability can emerge from underlying microscopic variability. Only a few historians have taken notice of this aspect of Planck's review, and even then only in passing. The point is mentioned briefly and without recognition of its conceptual weight.

Planck insisted on the generality of the method, its analytical precision, and the manner in which Gibbs defined temperature, entropy, and equilibrium entirely within the Hamiltonian framework. He also highlighted the elegant derivation of the canonical and microcanonical distributions. Planck examined with great care the possibility of assigning a fully mechanical meaning to temperature and entropy, and he noted the existence of several competing definitions. He also remarked on the insufficiency of purely Hamiltonian arguments for explaining irreversible processes, drawing attention to the role of indistinguishable particles and to the question of Poincaré recurrence.

Planck also pointed out a conceptual limitation in Gibbs's exposition: while the ensemble formalism assigns probabilities to regions of phase space, it contains no dynamical principle explaining why real systems should evolve toward the most probable macrostates. The canonical and microcanonical distributions are introduced as stationary measures justified by Liouville's theorem, but Gibbs does not account for the empirical fact that macroscopic systems spontaneously approach equilibrium. For Planck, this omission meant that Gibbs's theory—however rigorous in its treatment of equilibrium—did not provide a mechanical basis for irreversibility, central to the second law. Gibbs had characterised equilibrium with mathematical elegance, but not the process by which it is reached.

### *Stefan Meyer's Review*

In December 1903, Stefan Meyer devoted a review to Gibbs's book in the *Monatshefte für Mathematik und Physik*.[31] The journal itself had received a copy of the book for review. A specialist in radioactivity and a former student of Boltzmann, Meyer offered a thorough account, summarising and discussing all the essential chapters of the work.

The general tone is clearly positive, at times admiring: Meyer presents the book as a theoretical construction of exceptional rigour, establishing statistical mechanics on rational foundations. He identifies with precision what distinguishes Gibbs from Boltzmann: the abandonment of time-dependent arguments, the absence of any molecular chaos hypothesis, and the generality of a method independent of the nature

31 Meyer (1903).

of the system. In his view, the canonical ensemble constitutes the conceptual core of the theory.

These reviews establish that Gibbs's work was being read and discussed as early as the summer of 1902 in several major European scientific centres. In addition, Yale University's Gibbs Papers include a letter from Hendrik Antoon Lorentz to Gibbs dated April 1, 1902, written shortly after Lorentz received a copy of *Elementary Principles in Statistical Mechanics*. This document confirms that Lorentz was among the earliest European recipients of the treatise and provides direct archival evidence of its rapid circulation.[32]

Lorentz's subsequent engagement with Gibbs was not limited to private reading. Within a few years he began teaching Gibbs's ensemble-based statistical mechanics and incorporated it into published expositions based on his lecture notes. In particular, a 1907 theoretical-physics treatise of Lorentz derived from his teaching includes an extended summary of Gibbs's framework, and Lorentz later returned to Gibbs's method in a broader set of lectures delivered at the Collège de France and published in 1916. Such episodes indicate a comparatively rapid pedagogical assimilation of Gibbs's book in Leiden and its dissemination to wider scholarly audiences. This teaching context also shaped early research applications. Under Lorentz's supervision, Leonard S. Ornstein prepared and defended in 1908 a doctoral dissertation explicitly devoted to applying Gibbs's statistical mechanics to molecular-theoretical problems, illustrating how Gibbs's methods could be mobilised in concrete theoretical work.[33]

While our study focuses on the diffusion of Gibbs's book between 1902 and 1903, B. Popp, in a recent article, broadens the perspective by analysing a wider period of its reception.[34]

## Einstein's 1903 and 1904 Papers on Statistical Mechanics

After the first paper of 1902, Einstein published two further works on the same topic, in 1903 and 1904. Both studies develop the same conceptual programme: to derive the laws of thermodynamics from general mechanical and probabilistic principles, without relying on a specific kinetic or molecular model.

### *The Conceptual Extension of 1903*

Einstein's second paper on statistical mechanics, published in 1903 under the title "Eine Theorie der Grundlagen der Thermodynamik" ("A Theory of the Foundations of Thermodynamics"), extends the approach of the first one, but with a noticeably more abstract and formal tone, characteristic of a mathematical physicist.[35] Einstein generalises the dynamical framework by freeing himself from the strict Hamiltonian assumption. He considers general dynamical flows that preserve volume in phase space and for which the total energy is the only relevant integral of motion. He holds that, in order to establish thermodynamics, it is not at all necessary to assume that the equations of motion derive from a given Hamiltonian; it is enough to admit conservation of phase–space volume and conservation of energy. Hamiltonian mechanics becomes a special case of a much more general framework.

---

32 H. A. Lorentz to J. Willard Gibbs [Letter] (1902, Apr. 1), GEN MSS 1175, box 3, folder 87, Josiah Willard Gibbs Papers, Beinecke Rare Book and Manuscript Library, Yale University, New Haven, CT.
33 Popp (2024).
34 Popp (2024).
35 Einstein (1903).

This conceptual move marks a sharp break with Boltzmann. This paper is very far from any attempt to demonstrate the existence of atoms. Einstein does not seek to reconstruct thermodynamics from a particular microscopic dynamics, but to identify the minimal properties that guarantee the existence of a stationary statistical state depending only on the energy. He then introduces, in an implicit way, the essential hypothesis that the time average of a quantity for an isolated system coincides with the ensemble average—in other words, an undemonstrated ergodic hypothesis. He justifies this step by stating that the dynamical evolution leads the system to "explore" uniformly the region of phase space compatible with the energy, and therefore to be found most often in the most probable states. This idea, fundamental to his construction, is once again present in Gibbs's book. As Planck noted in his review, the identification between time averages and ensemble averages is founded neither mechanically nor mathematically, but rests on a practical postulate.

In this second article, Einstein becomes in a sense more "Gibbsian" than Gibbs: he pushes further the move to detach statistical mechanics from any particular mechanics, by insisting that the thermodynamic laws follow only from general properties of the dynamical flow—conservation of energy, volume invariance, and ergodicity—independently of any atomic interpretation or concrete mechanical model. Whereas Gibbs still relies on the Hamiltonian framework, Einstein allows himself to go beyond it. The contrast with Boltzmann becomes even more marked: Einstein sometimes borrows his vocabulary, but his conceptual architecture departs radically from Boltzmann's.

The 1903 paper is a formal construction with an unusually high level of abstraction for the literature of the time: it does not aim to justify thermodynamics microscopically, but to show that, under minimal structural conditions, the stationary form of the statistical distribution is unavoidable. It is a general probabilistic approach, freed from any explicit mechanics, in which the particular dynamics matters less than the existence of an invariant measure. What we see is not the approach of a disciple of Boltzmann, but the abstraction of a mathematical physicist who, in 1903, adopts a framework more general than that of Gibbs himself. This reading is corroborated by a private letter written by Einstein to Jakob Ehrat at the end of March 1903, at the very moment when the second paper was about to appear.[36] Referring explicitly to this work, Einstein describes it as a contribution to the foundations of thermodynamics. This formulation is highly revealing. Einstein does not present his

36 A. Einstein to J. Ehrat [Letter] (1903, last week of March), CPAE 5 English, doc. 11.

article as an attempt to establish the existence of atoms, nor as a kinetic derivation of thermodynamics, but as a foundational clarification of thermodynamics itself, based on minimal and deliberately non-mechanical assumptions.

### *The 1904 Paper*

In his 1904 paper, "On the General Molecular Theory of Heat," received by the *Annalen der Physik* in March, Einstein does not cite Gibbs's treatise—despite the fact that several reviews had appeared since 1902, most notably those by Max Abraham, Max Planck, and Stefan Meyer.[37] In particular, Planck's extensive review in the *Beiblätter* made the core structure of Gibbs's statistical mechanics accessible to a German-speaking readership well before Einstein submitted his paper. This context renders it plausible that Einstein was acquainted, at least indirectly, with the essential content of Gibbs's book by 1904.

A further point deserves attention. As noted above, Planck's 1903 review already contained a clear and explicit discussion of what we would now call statistical fluctuations. In two consecutive paragraphs, Planck analysed the deviations of mechanical quantities from their mean values and stressed that such deviations become negligible as the number of degrees of freedom increases. This provided a transparent physical explanation of why macroscopic systems display remarkably stable behaviour despite being governed by probabilistic laws. From this perspective, the conceptual framework for discussing fluctuations was already publicly available in the German literature several months before Einstein's publication.

Seen in this light, the appearance of fluctuations in Einstein's 1904 paper does not introduce a new concept. Rather, it represents a first and elementary application of a canonical result within a framework that Planck had recently articulated with notable clarity. Einstein's contribution consists in taking this general relation and illustrating its physical meaning through simple examples, rather than in developing new statistical machinery.

Einstein considers two schematic systems: a material gas and a volume of radiation. In both cases, he adopts the same minimal modelling assumptions, namely, that the system can be treated as composed of independent degrees of freedom and that the total energy is additive. Under these assumptions, fluctuations are directly linked to the mean energy and serve as a measure of what Einstein calls the "thermal stability" of the system. Large systems with many degrees of freedom are therefore highly stable, while small systems may exhibit perceptible fluctuations.

In the case of radiation, Einstein goes one step further by combining this statistical reasoning with well-established macroscopic thermodynamic relations. Drawing on the Stefan–Boltzmann law and Wien's displacement law, he reformulates the fluctuation criterion in terms of the volume of the cavity and a characteristic wavelength of the radiation. This allows him to express, in intuitive terms, the idea that fluctuations become relevant only when the system volume is comparable to the cube of a

37 Einstein (1904).

characteristic wavelength. For ordinary macroscopic volumes, the resulting estimate shows that relative energy fluctuations are extremely small, in agreement with the observed stability of thermal radiation at equilibrium. No new empirical prediction is introduced; the analysis serves to confirm the internal consistency of the statistical description.

Crucially, this reasoning remains entirely within a classical and continuous framework. Einstein does not introduce a microscopic model of radiation, does not rely on Planck's spectral law, and does not invoke any hypothesis of energy quanta. Wien's law functions purely as a phenomenological scaling relation linking temperature to a characteristic wavelength, not as a bridge to a deeper theory of radiation.

Several commentators have emphasised that the fluctuation relation used by Einstein in 1904 had already been derived by Gibbs. Pais, in particular, stresses that Einstein was unaware of this fact at the time.[38] Gearhart likewise points out that the energy fluctuation formula derived in Einstein's 1904 paper was already obtained within Gibbs's canonical framework.[39] More recently, Peliti and Rechtman explicitly recalled that Einstein's 1904 paper "obtains a result already derived by Gibbs," reinforcing the view that this aspect of the paper does not introduce a new formal element into statistical mechanics.[40]

While these assessments correctly identify the mathematical identity of the results, they are often embedded in retrospective narratives that emphasise the later importance of fluctuations in Einstein's subsequent work. From the perspective of the 1904 paper itself, however, the result functions as a straightforward application of canonical statistical mechanics, without any claim to formal novelty or foundational status.

## Correspondence with Paul Hertz (1910): Public Criticism and Private Discussions

In 1910, the physicist Paul Hertz (1881–1940) published two articles in the *Annalen der Physik* devoted to the mechanical foundations of thermodynamics.[41] At the time, Hertz was active as a *Privatdozent* in Heidelberg, where he had completed his habilitation in theoretical physics in 1908. His academic formation had taken place within the German university system, notably in Göttingen, where he received his doctorate in 1904. By 1910, Hertz thus occupied a recognised, though not senior, position within German theoretical physics.

In these two articles, Hertz offered a systematic examination of contemporary attempts to derive thermodynamics from mechanics. His discussion focused primarily on the works of Gibbs, Einstein, and Boltzmann, which he treated as the central reference points for modern statistical mechanics. The argumentative structure of the papers was organised around a sustained comparison between Gibbs's ensemble-based

38 Pais (1982).
39 Gearhart (1990).
40 Peliti & Rechtman (2017).
41 Hertz (1910a; 1910b).

framework and Einstein's statistical mechanics of 1902–1904, with Boltzmann occupying a more distant, largely historical position.

Hertz's general aim was not to deny the value of statistical methods, but to clarify their logical status. He argued that thermodynamic concepts such as entropy and irreversibility could not be fully grounded in deterministic mechanics alone and that probabilistic reasoning could not be introduced merely as a technical supplement. In this respect, his critique remained internal to the contemporary debate on the foundations of statistical mechanics and echoed earlier concerns raised in discussions of Boltzmann's programme.

The importance accorded to Hertz's intervention is underscored by its reception in Paul and Tatiana Ehrenfest's influential 1911 review article on the conceptual foundations of statistical mechanics. In contrast to their extremely brief treatment of Einstein's statistical papers of 1902–1903, the Ehrenfests engaged explicitly with Hertz's 1910 articles in the body of their analysis. Hertz's arguments were discussed as part of the contemporary debate on the logical status of statistical ensembles and the justification of probabilistic reasoning in mechanics. Even where the Ehrenfests did not fully endorse his conclusions, they treated his critique as raising genuine conceptual difficulties requiring clarification. This asymmetry is revealing: whereas Einstein's early statistical work was classified as compatible with an already established Gibbsian framework and therefore required little comment, Hertz's critique was perceived as reopening foundational questions.[42]

The historiographical significance of Hertz's intervention lies less in its technical conclusions than in its framing. The first article opens with the striking assertion: "It is hardly possible to find a fundamental idea or a basic formula in these questions that was not already contained in the work of Gibbs."[43] Although no specific author is named in this sentence, the surrounding analysis makes clear that Hertz was addressing mechanical and quasi-mechanical approaches to thermodynamics prominent in the German-speaking literature. Within this landscape, Einstein's statistical trilogy occupied a privileged position: it was analysed in detail and repeatedly juxtaposed with Gibbs's ensemble theory. The remark thus amounted to an explicit public recognition of a close structural parallel between Einstein's statistical mechanics and Gibbs's work.

Einstein reacted promptly to Hertz's critique, but in a revealing manner. In August 1910, he entered into a private correspondence with Hertz explicitly to avoid a public controversy in the *Annalen der Physik*.[44] In a letter dated August 14, 1910, Einstein expressed his appreciation of Hertz's paper and proposed that points of disagreement be discussed in person, adding that a printed debate would be undesirable. At that stage, Einstein indicated that he intended to remain in Zurich until September and suggested that Hertz might visit him there.

42 Ehrenfest & Ehrenfest (1911).
43 Hertz (1910a).
44 A. Einstein to P. Hertz [Letter] (1910, Aug. 14), CPAE 5 English, doc. 220.

A second letter, dated August 26, 1910, shows that Einstein instead agreed to meet Hertz in Basel, despite the fact that no such journey had originally been planned.[45] Einstein arranged accommodation for himself and coordinated the meeting in detail, even proposing to meet Hertz at the railway station. This arrangement and the associated logistical effort indicate that Einstein regarded the matter as sufficiently important to warrant a carefully coordinated private meeting.

Hertz's letters to Einstein have not been preserved. Nevertheless, the subsequent public record is suggestive. In a short note published in 1911 in the *Annalen der Physik*, Einstein acknowledged the force of Hertz's critique and stated that acquaintance with Gibbs's treatise would have altered his own publication strategy. He declared that, had he known Gibbs's work at the time, he would not have published the statistical papers of 1902–1904.[46]

The two surviving letters from Einstein to Hertz, dated August 14 and 26, 1910, are published with a brief editorial annotation in the *Collected Papers of Albert Einstein*. To our knowledge, they have not been examined in the secondary literature as evidence for a deliberate strategy aimed at managing the public framing of Einstein's early work in statistical mechanics.

Taken together, Hertz's critique of 1910, Einstein's preference for a private exchange culminating in a meeting in Basel, and the restrained public statement of 1911 mark a significant moment in the management of Einstein's early statistical work. This episode shows that the close parallel between Einstein's statistical mechanics and Gibbs's framework was recognised by informed contemporaries and that Einstein ultimately accepted this assessment in print, without attempting to defend a claim of methodological or conceptual independence.

## Einstein in 1949: Retrospective Account of the Genesis of His Work on Statistical Mechanics

In his *Autobiographical Notes*, Einstein writes: “Not knowing the previous investigations of Boltzmann and Gibbs, which had already been published and in fact had exhausted the subject, I developed statistical mechanics and the molecular kinetic theory of thermodynamics from the mechanical equations.”[47] This often-quoted sentence is the principal explicit mention of Gibbs in Einstein's autobiographical writings. It retrospectively frames Einstein's early work on statistical mechanics (1902–1904) as having been carried out without prior knowledge of either Boltzmann's or Gibbs's contributions, and thus as an independent reconstruction grounded solely in mechanical equations and probabilistic reasoning.

Einstein further adds that his main objective at the time was “to find facts that would guarantee as much as possible the existence of atoms of definite size.” He

45 A. Einstein to P. Hertz [Letter] (1910, Aug. 26), CPAE 5 English, doc. 222.
46 Einstein (1911).
47 Einstein (1949).

contrasts this aim with what he presents as Gibbs's programme, namely the establishment of general regularities in systems with many degrees of freedom. This atomistic and goal-oriented framing, however, sits uneasily with the actual content of Einstein's papers of 1902–1904. These texts are predominantly formal in character: they develop equilibrium distributions and canonical weights using very general invariance principles—most notably Liouville's theorem—without detailed modelling of particles, trajectories, or collision-based kinetic theory. Explicit analyses of fluctuations are absent from the papers of 1902 and 1903 and appear only in the 1904 article.

Einstein's 1949 statements therefore do not align well with the internal structure or declared aims of his early statistical papers. Rather than constituting an experimental or phenomenological programme directed toward the proof of atomism, these works are best read as a theoretical reconstruction of thermodynamics within a general probabilistic framework.

Einstein's autobiographical remarks cannot be treated as straightforward historical reports. Well before 1902, Einstein was demonstrably familiar with Boltzmann's work: his correspondence with Mileva Marić contains repeated and enthusiastic references to Boltzmann's writings, in particular to the *Vorlesungen über Gastheorie*, which he read closely and discussed at length.[48] The claim, made retrospectively in 1949, that he developed statistical mechanics without knowledge of Boltzmann is therefore incompatible with the documentary record.

With respect to Gibbs, however, no definitive conclusion can be drawn from Einstein's retrospective statements alone. The assertion of non-knowledge rests largely on autobiographical testimony and on the absence of explicit citations, both of which require cautious interpretation.

Einstein's 1949 account is thus best understood as a retrospective reconstruction shaped by his later methodological views and by a self-image emphasising independence and conceptual autonomy. As such, it cannot serve as decisive evidence for the historical conditions under which his work of 1902–1904 was produced. The question of Einstein's possible acquaintance with Gibbs's treatise must instead be addressed through an examination of the material circulation, accessibility, and contemporary reception of Gibbs's book—an analysis to which the present study is devoted.

## Possible Channels of Access to Gibbs's Treatise in April 1902

Two distinct channels may have enabled Einstein to gain early access to Gibbs's *Elementary Principles* during the spring of 1902.

The first is institutional. As shown above, the University Library of Basel received its copy of Gibbs's book on April 10, 1902 as part of the *Yale Bicentennial Publications*. Newly acquired scientific monographs were not necessarily available for immediate loan, but they could normally be consulted on site once received and processed by

48 Gearhart (1990).

the library. Within this institutional route, members of Einstein's immediate circle may have played a mediating role. His close friend Conrad Habicht, registered since May 23, 1901, as a doctoral student in mathematics at the University of Bern, belonged to a mobile network of young Swiss mathematicians circulating between Zürich, Bern, and Basel. His dissertation, devoted to Steiner's circle series, placed him within a classical geometrical tradition connected to the broader Swiss mathematical community. In this context, it would have been natural for Habicht to consult the University Library of Basel—whose mathematical holdings were richer than those of Bern—while preparing his doctoral work. Should he have visited Basel in April or May 1902, he could have encountered the newly arrived Gibbs volume on display and mentioned its existence or contents to Einstein upon returning to Bern. No documentary evidence attests such a visit or consultation, and this scenario therefore remains conjectural.

The second channel is individual. Dr. Max Talmey, who visited Einstein in Bern in April 1902, had been living and working in New York, where Gibbs's treatise had just been published by Charles Scribner's Sons. Talmey had played a formative role in Einstein's intellectual development during the 1890s by supplying him with scientific and philosophical literature, and he remained, by his own account, an eclectic and enthusiastic reader. No testimony confirms that he carried or mentioned Gibbs's book during his visit. Nonetheless, the close coincidence of dates makes this possibility worth noting, even if it appears less probable than the institutional route.

Neither of these hypotheses—one institutional (possibly mediated through Habicht), the other personal—can be confirmed on the basis of the surviving evidence. Their importance lies elsewhere. Taken together, they show that Einstein's access to Gibbs's *Elementary Principles* in April 1902, long regarded as highly unlikely, was in fact materially plausible.

## Historiographical Implications of a Possible Early Access to Gibbs (1902)

The findings presented in this article have implications that extend beyond questions of priority or technical parallelism in early 20th-century statistical mechanics. They directly affect a long-standing historiographical interpretation according to which Albert Einstein could not have known Gibbs's *Elementary Principles in Statistical Mechanics* when writing his first paper on statistical mechanics in 1902, and that his work must therefore be regarded as having developed independently of Gibbs's framework.

This interpretation rested primarily on biographical and testimonial considerations. Einstein's marginal institutional position in 1902—outside the university system and without stable academic affiliation—was taken to make access to newly published specialist literature unlikely. In addition, Einstein himself later stated that he had not known Gibbs's book at the time and even suggested that such knowledge would have altered his decision to publish his early statistical papers. These retrospective statements, combined with his biographical situation, led historians to treat non-access to Gibbs as a settled assumption rather than as an open historical question.

The archival reconstruction presented here modifies this background condition. The documented arrival of the 1902 Gibbs volume in several European university libraries in early spring 1902, including Basel on April 10, establishes that the book was materially present in Einstein's immediate scientific environment more than 2 months before the submission of his first statistical paper on June 26, 1902. While this does not demonstrate that Einstein actually consulted the book, it removes the implicit argument of material impossibility that has tacitly underpinned claims of non-access.

Once this material context is restored, the historiographical situation changes in a significant way. The absence of explicit references to Gibbs in Einstein's early papers can no longer be treated as straightforward evidence of ignorance, and retrospective self-testimonies must be interpreted with greater methodological caution. The historical problem shifts from one of near-impossibility to one of plausibility: access becomes a question to be assessed rather than an assumption to be presupposed.

This shift has consequences for the interpretation of Einstein's early statistical work. His papers of 1902–1904 can still be read as autonomous theoretical constructions, but they can no longer be understood as having been produced in a documentary vacuum. They were written at a moment when a comprehensive and systematic framework for statistical mechanics had just become available in Europe, and this contextual availability must now be taken into account when assessing their intellectual situation.

In the spring of 1902, Einstein was in a professionally and intellectually unstable position, without a permanent post and still in the process of defining his scientific orientation. In such a context, access to a work like Gibbs's *Elementary Principles*, which presented statistical mechanics as a unified and mathematically rigorous foundation for thermodynamics and molecular physics, could have exerted a significant structuring influence. This does not imply direct borrowing or textual dependence, but it does suggest that Einstein's engagement with statistical mechanics took place against the background of an already articulated theoretical synthesis.

From this perspective, the years 1902–1904 may be understood as a period of conceptual consolidation rather than as the outcome of an isolated and self-contained development. The originality of Einstein's later contributions, particularly those of 1905, would then lie not in the abrupt emergence of statistical ideas from isolation, but in their critical transformation and physical reinterpretation. Statistical concepts were mobilised as tools for probing physical reality, most notably through the attribution of physical and experimental significance to fluctuations.

Such an interpretation does not rest on biographical speculation, nor does it attribute specific results or methods directly to Gibbs. Rather, it highlights a general historiographical point: access to a major theoretical synthesis at a formative moment can play an important role in shaping the conceptual landscape within which new problems are formulated and explored. Restoring the material availability of Gibbs's book therefore invites a reassessment of the narrative of independence that has long structured accounts of Einstein's early statistical mechanics.

More broadly, the possibility of an early encounter with Gibbs's *Elementary Principles* has implications for the extensive historiography devoted to Einstein's scientific development between 1901 and 1905. Much of this literature rests on the assumption that Einstein's engagement with statistical mechanics, and his subsequent breakthroughs, emerged in relative isolation and without the guidance of an existing systematic framework. Once the material availability of Gibbs's book in Einstein's environment is acknowledged, this assumption can no longer be taken for granted. The years 1901–1905 may instead be reinterpreted as a period during which Einstein was exposed to, and potentially assimilated, one of the most advanced syntheses of statistical mechanics available at the time. Such an encounter could have played an important role in shaping his scientific orientation, providing a conceptual horizon against which problems were selected, reformulated, and ultimately transformed. Recognising this possibility invites a reassessment of the narrative that portrays Einstein's early career as a sequence of independent breakthroughs, and encourages a more integrated view of his formative years as a process of conceptual appropriation, consolidation, and critical reworking within an already structured theoretical landscape.

## Theoretical Physics as a Solitary Enterprise According to Einstein

The historiographical reassessment proposed in this article also invites a broader reflection on Einstein's own representation of theoretical work and on the public image of the theoretical physicist that he actively helped to shape. In several retrospective and public writings, Einstein portrayed theoretical physics as an activity that is largely solitary in character, carried out at some distance from academic routines, collective research programmes, and institutional frameworks. This representation placed particular emphasis on inward intellectual work, conceptual synthesis, and intuition, rather than on cumulative traditions or collaborative practices.

The purpose of this section is not to reopen questions of priority in other fields, but to identify a recurrent mode of retrospective self-representation that bears directly on the interpretation of Einstein's later statements about Gibbs.

A particularly clear expression of this view is found in the preface Einstein wrote in 1933 for Max Planck's *Where Is Science Going?*, where he declared:

> The supreme task of the physicist is to arrive at those universal elementary laws from which the cosmos can be built up by pure deduction. There is no logical path to these laws; only intuition, resting on sympathetic understanding of experience, can reach them.[49]

---

49 Einstein (1933).

In this formulation, theoretical discovery appears as an essentially individual and inward activity. The emphasis on intuition and on the absence of a "logical path" implicitly downplays the role of shared problem spaces, collective traditions, and cumulative research practices in favour of singular intellectual insight.

This normative conception of theoretical physics is closely connected to a broader feature of Einstein's scientific writings during his most productive period, roughly from 1902 to 1917: a striking reticence in acknowledging sources, influences, and prior work. The absence of references in Einstein's 1905 paper on special relativity is a well-known example. That article contains no explicit discussion of the extensive theoretical and experimental literature that preceded it, including the work of Lorentz, Poincaré, Larmor, Michelson and Morley, or Trouton and Noble. A decade of intense developments in electrodynamics, electron theory, and ether-drift experiments is effectively absent from the text.

This narrative pattern is not confined to the paper of 1905, but recurs across different genres of writing. In the popular brochure *On the Special and the General Theory of Relativity*, explicitly intended for a non-specialist audience, the same de-historicised presentation is adopted.[50] Lorentz appears only as a nominal reference attached to the transformations that bear his name, without any discussion of the physical motivations, heuristic struggles, or conceptual context that led him to introduce them. The transformations are presented as formal relations required by the theory, not as the endpoint of a long-standing research programme in electrodynamics. Henri Poincaré, whose work addressed both the relativity principle and the mathematical structure of these transformations, is entirely absent from the exposition. The pedagogical aim of the text thus does not lead to greater historical contextualisation; on the contrary, it reinforces a narrative in which conceptual necessity replaces historical development.

The same logic governs Einstein's public intervention in the international press following the 1919 eclipse expeditions. In his article "Time, Space, and Gravitation," published in *The Times*, he again reconstructs special relativity as emerging almost seamlessly from the internal tensions of Maxwellian electrodynamics.[51] The constancy of the velocity of light and the principle of relativity are presented as empirically grounded constraints that compel a revision of kinematics. Here too, Lorentz's name functions primarily as a label for a mathematical transformation, detached from its historical genesis, while Poincaré remains entirely unmentioned. The continuity between Newtonian mechanics and Einstein's theory is emphasised, whereas the complex transitional landscape of late 19th- and early 20th-century electrodynamics is largely effaced.

This pattern persists in Einstein's later methodological reflections. In his 1940 address "Considerations Concerning the Fundaments of Theoretical Physics," published in *Science*, Einstein offered a retrospective account of the emergence of special relativity, in which the theory appears to arise almost naturally from the internal

50 Einstein (1917).
51 Einstein (1919).

structure of Maxwellian electrodynamics.[52] The empirical constancy of the velocity of light leads to the Lorentz transformations, which in turn impose a new kinematical framework. The transition is depicted as a consequence of formal and empirical constraints rather than as the outcome of a historically situated research programme. Lorentz is mentioned only in connection with his incorporation of the electron into electrodynamics, and in highly laudatory terms; his extensive pre-1905 work on transformation theory and local time plays no role in this reconstruction, while Poincaré remains entirely absent.

Even in explicitly commemorative contexts, the same selective attribution can be observed. In his 1953 tribute marking the centenary of Hendrik A. Lorentz's birth, Einstein celebrated Lorentz as a towering figure of 19th-century physics and as a master of theoretical refinement. Yet Lorentz's contribution to the conceptual and mathematical framework that directly preceded special relativity was once again left unaddressed.[53] The transition to special relativity was not narrated as a historically continuous development, but as a conceptual reorganisation driven by internal necessities.

The long-term effects of this narrative construction extend well beyond Einstein's own writings. For more than a century, special relativity has been taught largely through the conceptual and rhetorical framework established by Einstein himself. In the vast majority of university textbooks, Lorentz appears primarily as an eponym attached to a set of coordinate and time transformations, while his broader physical programme, heuristic motivations, and sustained engagement with the foundations of electrodynamics are largely absent. The historical role of Lorentz is thus reduced to a formal label, and the complex prehistory of special relativity is compressed into a minimal backdrop against which Einstein's theory appears as an almost inevitable conceptual resolution. This pedagogical canonisation has reinforced a de-historicised understanding of the theory's origins and has contributed to the enduring marginalisation of alternative narratives emphasising continuity, collective development, and the cumulative structure of theoretical physics at the turn of the 20th century.

A similar pattern can be identified in Einstein's work on molecular physics. In his 1905 paper on Brownian motion and in his doctoral thesis on molecular dimensions, Einstein made no reference to the work of William Sutherland (1859–1911), despite documentary evidence that Sutherland's ideas were discussed in his immediate intellectual environment. In particular, the correspondence between Einstein and Michele Besso in 1903 explicitly refers to Sutherland's hypothesis in connection with molecular dimensions, viscosity, diffusion, and the applicability of hydrodynamic laws at the molecular scale.[54] Sutherland had developed a research programme linking molecular forces, viscosity, diffusion, and osmotic pressure, with the aim of using macroscopic transport phenomena as probes of molecular size and mass. Several elements of

52 Einstein (1940).
53 Einstein (1953).
54 M. Besso to A. Einstein [Letter] (1903, Feb. 7–11), CPAE 5 English, doc. 6; A. Einstein to M. Besso [Letter] (1903, Mar. 17), CPAE 5 English, doc. 7.

this programme resonate with Einstein's 1905 treatment of molecular dimensions and Brownian motion, although Einstein's own contribution lay in the statistical reformulation of these problems and in their connection to experimentally accessible quantities. Here again, a relevant prior programme remained unacknowledged in the published papers.

Even in statistical mechanics, where Boltzmann's influence has long been emphasised by historians, explicit references are targeted. In his 1902 paper, Einstein cites specific results of Boltzmann without explicitly attributing the overall construction to him. Once again, later reconstructions have inferred intellectual debts beyond those explicitly acknowledged in print.

These cases suggest that the silence surrounding sources and influences in Einstein's early work was neither accidental nor confined to a single episode. Rather, it reflects a consistent mode of presentation in which theoretical advances were framed as emerging primarily from individual reasoning and intuition. This narrative strategy does not accurately reflect Einstein's actual scientific practice, which remained embedded in dense networks of reading, correspondence, and discussion. Instead, it functioned as a symbolic resource in the construction of a public image of theoretical physics as a solitary enterprise.

From this perspective, Einstein's retrospective statements concerning his lack of familiarity with Gibbs's book in 1902 appear less as straightforward factual reports than as elements of a broader narrative pattern. The case of Gibbs thus gains historiographical significance not because it establishes a direct influence, but because it exposes the limitations of relying on silence and self-testimony as indicators of intellectual independence. More generally, it highlights the need to distinguish between the collective conditions under which scientific work is produced and the individualised narratives through which it is later presented and remembered.

***Acknowledgements***

I am grateful to the anonymous reviewers for their careful reading of the manuscript and for their constructive comments and suggestions. I also thank the editors for their guidance and support throughout the review process.

## References

Abraham, M. (1902). [Review of J. Willard Gibbs, *Elementary principles in statistical mechanics: Developed with especial reference to the rational foundation of thermodynamics* (New York: Charles Scribner's Sons, 1902)]. *Physikalische Zeitschrift, 3*(24), 582–584.

Born, M. (1956). *Physics in my generation*. London, UK: Pergamon Press.

Bryan, G. H. (1902). [Review of J. Willard Gibbs, *Elementary principles in statistical mechanics*]. *Nature, 66*, 291–292.

Darrigol, O. (2004). The origins of the entropy concept. In J. Dalibard, B. Duplantier, & V. Rivasseau (Eds.), *Poincaré seminar 2003: Bose–Einstein condensation—Entropy* (pp. 101–118). Basel, Switzerland: Birkhäuser.

Darrigol, O., & Renn, J. (2013). The emergence of statistical mechanics. In J. Z. Buchwald & R. Fox (Eds.), *The Oxford handbook of the history of physics* (pp. 765–788). Oxford, UK: Oxford University Press.

Duncan, A., & Janssen, M. (2019). *Constructing quantum mechanics, Volume 1: The scaffold 1900–1923*. Oxford, UK: Oxford University Press.

Ehrenfest, P., & Ehrenfest, T. (1911). Begriffliche Grundlagen der statistischen Auffassung in der Mechanik. In F. Klein & C. Müller (Eds.), *Encyklopädie der mathematischen Wissenschaften mit Einschluss ihrer Anwendungen* (Vol. 4, Part 4, Article 32, pp. 1–90). Leipzig, Germany: B. G. Teubner.

Einstein, A. (1901). Folgerungen aus den Capillaritätserscheinungen. *Annalen der Physik, 4*, 513–523.

Einstein, A. (1902a). Kinetische Theorie des Wärmegleichgewichtes und des zweiten Hauptsatzes der Thermodynamik. *Annalen der Physik, 9*, 417–433.

Einstein, A. (1902b). Über die thermodynamische Theorie der Potentialdifferenz zwischen Metallen und vollständig dissociierten Lösungen ihrer Salze und über eine elektrische Methode zur Erforschung der Molekularkräfte. *Annalen der Physik, 8*, 798–814.

Einstein, A. (1903). Eine Theorie der Grundlagen der Thermodynamik. *Annalen der Physik, 11*, 170–187.

Einstein, A. (1904). Zur allgemeinen molekularen Theorie der Wärme. *Annalen der Physik, 14*, 354–362.

Einstein, A. (1911). Bemerkungen zu den P. Hertzschen Arbeiten: Über die mechanischen Grundlagen der Thermodynamik. *Annalen der Physik, 34*, 175–176.

Einstein, A. (1917). *Über die spezielle und die allgemeine Relativitätstheorie (gemeinverständlich)*. Braunschweig, Germany: Friedrich Vieweg & Sohn.

Einstein, A. (1919, November 28). Time, space, and gravitation. *The Times* (London), pp. 13–14.

Einstein, A. (1933). Preface. In M. Planck, *Where is science going?* (pp. 9–14). London, UK: Allen & Unwin.

Einstein, A. (1940). Considerations concerning the fundaments of theoretical physics. *Science, 91*(2369), 487–492. https://doi.org/10.1126/science.91.2369.487

Einstein, A. (1949). Autobiographical notes. In P. A. Schilpp (Ed.), *Albert Einstein: Philosopher–scientist* (Vol. 7, pp. 1–94). La Salle, IL: Open Court.

Einstein, A. (1953). *H. A. Lorentz: His creative genius and his personality* (Communication from the National Museum for the History of Science, Leyden, no. 91). Leiden, The Netherlands: Rijksmuseum voor de Geschiedenis der Natuurwetenschappen.

Gearhart, C. A. (1990). Einstein before 1905: The early papers on statistical mechanics. *American Journal of Physics, 58*, 468–480.

Giacomini, H. (2026). The rapid arrival of Josiah Willard Gibbs's *Elementary principles in statistical mechanics* in European university libraries. *Archive for History of Exact Sciences, 80*(1), Article 4. https://doi.org/10.1007/s00407-026-00362-2

Gibbs, J. W. (1902). *Elementary principles in statistical mechanics*. New York, NY: Charles Scribner's Sons.

Gordin, M. D. (2022). Einsteinian language: Max Talmey, Benjamin Lee Whorf and linguistic relativity. *The British Journal for the History of Science, 55*, 145–165.

Gutzmer, A. (1902). VI. Mitteilungen und Nachrichten. 1. Akademien. Gesellschaften. Vereinigungen. Versammlungen. *Jahresbericht der Deutschen Mathematiker-Vereinigung, 11*, 302.

Hertz, P. (1910a). Über die mechanischen Grundlagen der Thermodynamik. Erste Abhandlung. *Annalen der Physik, 33*, 225–274.

Hertz, P. (1910b). Über die mechanischen Grundlagen der Thermodynamik. Zweite Abhandlung. *Annalen der Physik, 33*, 537–552.

Janssen, M., & Lehner, C. (Eds.). (2014). *The Cambridge companion to Einstein*. Cambridge, UK: Cambridge University Press.

Klein, M. J. (1963). Einstein's first paper on quanta. *The Natural Philosopher, 2*, 59–86.

Klein, M. J. (1967). Thermodynamics in Einstein's thought. *Science, 157*, 509–516.

Klein, M. J. (1987). Some historical remarks on the statistical mechanics of Josiah Willard Gibbs. In J. L. Berggren & B. R. Goldstein (Eds.), *From ancient omens to statistical mechanics: Essays on the exact sciences presented to Asger Aaboe* (pp. 281–289). Copenhagen, Denmark: University Library.

Klein, M. J., Kox, A. J., Renn, J., & Schulmann, R. (Eds.). (1995). *The collected papers of Albert Einstein, Vol. 5: The Swiss years: Correspondence, 1902–1914. English translation supplement* (A. Beck, Trans.; D. Howard, Consultant). Princeton, NJ: Princeton University Press.

Kox, A. J. (2014). Einstein on statistical physics: Fluctuations and atomism. In M. Janssen & C. Lehner (Eds.), *The Cambridge companion to Einstein* (pp. 103–116). Cambridge, UK: Cambridge University Press.

Mehra, J. (1975). Einstein and the foundation of statistical mechanics. *Physica A, 79*, 447–477.

Meyer, S. (1903). [Review of J. Willard Gibbs, *Elementary Principles in Statistical Mechanics*]. *Monatshefte für Mathematik und Physik, 14*, A55–A59.

Navarro, L. (1998). Gibbs, Einstein and the foundations of statistical mechanics. *Archive for History of Exact Sciences, 53*(2), 147–180.

Pais, A. (1982). *"Subtle is the lord…."* Oxford, UK: Oxford University Press.

Peliti, L., & Rechtman, R. (2017). Einstein's approach to statistical mechanics: The 1902–1904 papers. *Journal of Statistical Physics, 167*, 1020–1038.

Planck, M. (1903). [Review of J. W. Gibbs: *Elementary principles in statistical mechanics*]. *Beiblätter zu den Annalen der Physik, 27*, 748–753.

Popp, B. D. (2024). *Contemporary reaction to Gibbs's statistical mechanics*. https://doi.org/10.48550/arXiv.2412.11425

Renn, J. (2005). Einstein's invention of Brownian motion. *Annalen der Physik, 14*(S1), 23–37.

Stachel, J., Cassidy, D. C., Renn, J., & Schulmann, R. (Eds.). (1987a). *The collected papers of Albert Einstein, Vol. 1: The early years, 1879–1902*. Princeton, NJ: Princeton University Press.

Stachel, J., Cassidy, D. C., Renn, J., & Schulmann, R. (Eds.). (1987b). *The collected papers of Albert Einstein, Vol. 1: The early years, 1879–1902. English translation supplement* (A. Beck, Trans.). Princeton, NJ: Princeton University Press.

Stachel, J., Cassidy, D. C., Renn, J., & Schulmann, R. (1989). Editorial note: Einstein on the foundations of statistical physics. In J. Stachel (Ed.), *The collected papers of Albert Einstein, Vol. 2: The Swiss years: Writings, 1900–1909* (pp. 41–55). Princeton, NJ: Princeton University Press.

Talmey, M. (1932). *The relativity theory simplified and the formative period of its inventor*. New York, NY: Falcon Press.

Tolman, R. C. (1938). *The principles of statistical mechanics*. Oxford, UK: Clarendon Press.